\documentclass[prx,twocolumn,superscriptaddress,floatfix,nopacs]{revtex4-1}
\usepackage{graphicx,amsfonts,amssymb,amsmath,hyperref,enumerate}

\newif\ifhyper
\hypertrue
\ifhyper
\hypersetup{
   citecolor = {green},
   colorlinks = {true}, 
   urlcolor = {blue} 
}

\newcommand{\w}{{\omega}}

\begin{document} 

\title{Quantum Portfolio Optimization with Investment Bands and Target Volatility}

\author{Samuel Palmer}
\affiliation{Multiverse Computing, Centre for Social Innovation, 192 Spadina Ave, Suite 412, Toronto M5T 2C2, Canada}

\author{Serkan Sahin}
\affiliation{Multiverse Computing, Paseo de Miram\'on 170, E-20014 San Sebasti\'an, Spain}

\author{Rodrigo Hern\'andez}
\affiliation{Multiverse Computing, Paseo de Miram\'on 170, E-20014 San Sebasti\'an, Spain}

\author{Samuel Mugel}
\affiliation{Multiverse Computing, Centre for Social Innovation, 192 Spadina Ave, Suite 412, Toronto M5T 2C2, Canada}

\author{Rom\'an Or\'us}
\affiliation{Donostia International Physics Center, Paseo Manuel de Lardizabal 4, E-20018 San Sebasti\'an, Spain}
\affiliation{Ikerbasque Foundation for Science, Maria Diaz de Haro 3, E-48013 Bilbao, Spain}
\affiliation{Multiverse Computing, Paseo de Miram\'on 170, 20014 San Sebasti\'an, Spain}

\begin{abstract}
In this paper we show how to implement in a simple way some complex real-life constraints on the portfolio optimization problem, so that it becomes amenable to quantum optimization algorithms. Specifically, first we explain how to obtain the best investment portfolio with a given target risk. This is important in order to produce portfolios with different risk profiles, as typically offered by financial institutions. Second, we show how to implement individual investment bands, i.e., minimum and maximum possible investments for each asset. This is also important in order to impose diversification and avoid corner solutions. Quite remarkably, we show how to build the constrained cost function as a quadratic binary optimization (QUBO) problem, this being the natural input of quantum annealers. The validity of our implementation is proven by finding the optimal portfolios, using D-Wave Hybrid and its \emph{Advantage} quantum processor, on portfolios built with all the assets from S$\&$P100 and S$\&$P500. Our results show how practical daily constraints found in quantitative finance can be implemented in a simple way in current NISQ quantum processors, with real data, and under realistic market conditions. In combination with clustering algorithms, our methods would allow to replicate the behaviour of more complex indexes, such as Nasdaq Composite or others, in turn being particularly useful to build and replicate Exchange Traded Funds (ETF). 

\end{abstract}

\maketitle

\emph{Introduction.---} The problem of portfolio optimization deals with maximizing the return and minimizing the risk of the investment in a set of assets \cite{portfolio}. As simple as it sounds, this is the most paradigmatic example of an optimization problem in quantitative finance. Its importance is clear, since it is at the core of many financial objects that affect our daily lives: pension plans, ETFs, investment funds, and many more. As such, the problem is well-known to be computationally intractable in realistic settings. It is well known that if investments come in discrete units, then the problem becomes NP-Hard even if investments are done for a single trading step. The problem becomes even harder if we search for optimal trading trajectories in a period of time, where one should include further constraints such as transaction costs and market impact. But even in the static case, brokers building actual portfolios in real life handle lots of complex constraints that make the problem quickly intractable, with almost no resemblance to academic toy models. In this scenario, quantum computing has come up as a promising tool to handle intractable financial problems \cite{Orus2018}. Optimizing portfolios has been one of the first applications of quantum annealers \cite{delprado, Rosenberg2016}. Recent calculations with real data \cite{Mugel2020, Mugel2020b} have shown that hybrid quantum annealing is a good approach to handle this problem.  

Here we show how two important constraints in portfolio optimization, arising in the daily life of a quantitative analyst, can be implemented in a language that is natural for quantum computers, and in particular for quantum annealers. To be specific, we first explain \emph{ how to target optimal investment portfolios with a fixed volatility.} This is very useful, since financial marketplaces always offer a wide spectrum of products for clients with different risk profiles: conservative, medium-risk, high-risk, and so on. Second, we also show here \emph{how to impose investment bands in the computed portfolios}. This means that the investment for each asset is between a minimum and a maximum, this being interval asset-dependent and/or sector-dependent. We will see that all these constraints can be implemented naturally with a cost function that amounts to a Quadratic Unconstrained Binary Optimization (QUBO) problem. Finally, we prove the validity of our implementation by computing the optimal portfolios for investments in all the assets of S$\&$P100 and S$\&$P500, using D-Wave's hybrid quantum annealing algorithm with the \emph{Advantage} processor. To the best of our knowledge, this calculation is the most realistic static portfolio optimization carried so far on a quantum computer. 

\emph{The model.---} As we have already introduced in previous papers (see for instance Refs.\cite{Mugel2020, Mugel2020b}), and according to Modern Portfolio Theory, the optimal investment at a defined level of risk is the one which maximizes profit \cite{Singleton2018}. The risk taken by the investor in the portfolio is measured by the volatility $\sigma$, which is computed from the covariance matrix $\Sigma$ as 
\begin{equation}
\sigma \equiv \sqrt{\w^T \Sigma \w},  
\end{equation}
with $\w \equiv \bar{\w}/K$ a vector of components $\w_n \in [ 0, 1]$, being these the fraction of the total investment in asset $n = 1, 2, \cdots, N$. Here $N$ is the total number of assets, $K$ the total amount invested, and $\bar{\w}$ the vector of actual investments in each asset. The optimal portfolio is then the one that minimizes the cost function 
\begin{equation}
\label{eq:bare_cost_function}
H =
 -\mu^T \w
 + \frac{\gamma}{2} \w^T \Sigma \w.
\end{equation}
In the above equation, $\mu$ is the vector of logarithmic returns and parameter $\gamma$ is the so-called \emph{risk aversion}, which controls the portfolio's penalty for risk, i.e., the amount of risk an investor is willing to take. Both the logarithmic returns and covariance matrix can be computed straightforwardly from the stocks' values, see for instance Ref. \cite{Mugel2020}. In practice, we would also like to fix the entire budget being invested to $K$. In this formalism, this is guaranteed by enforcing the normalization of holdings $\w$ via a Lagrange multiplier $\rho$, so that $\sum_n \w_{n} = 1$. The cost function is then given by  
\begin{equation}
H = -\mu^T \w
 + \frac{\gamma}{2} \w^T \Sigma \w + \rho \left(\sum_n \w_{n} - 1 \right)^2.
\end{equation}
Furthermore, we also assume that  shares can only be sold in large bundles. These constraints imply that our objective variables $\w_n$ are integer variables and, therefore, we are dealing with the discrete version of the portfolio optimization problem, which is NP-Hard. 

\begin{figure}
	\centering
	\includegraphics[width=1\linewidth]{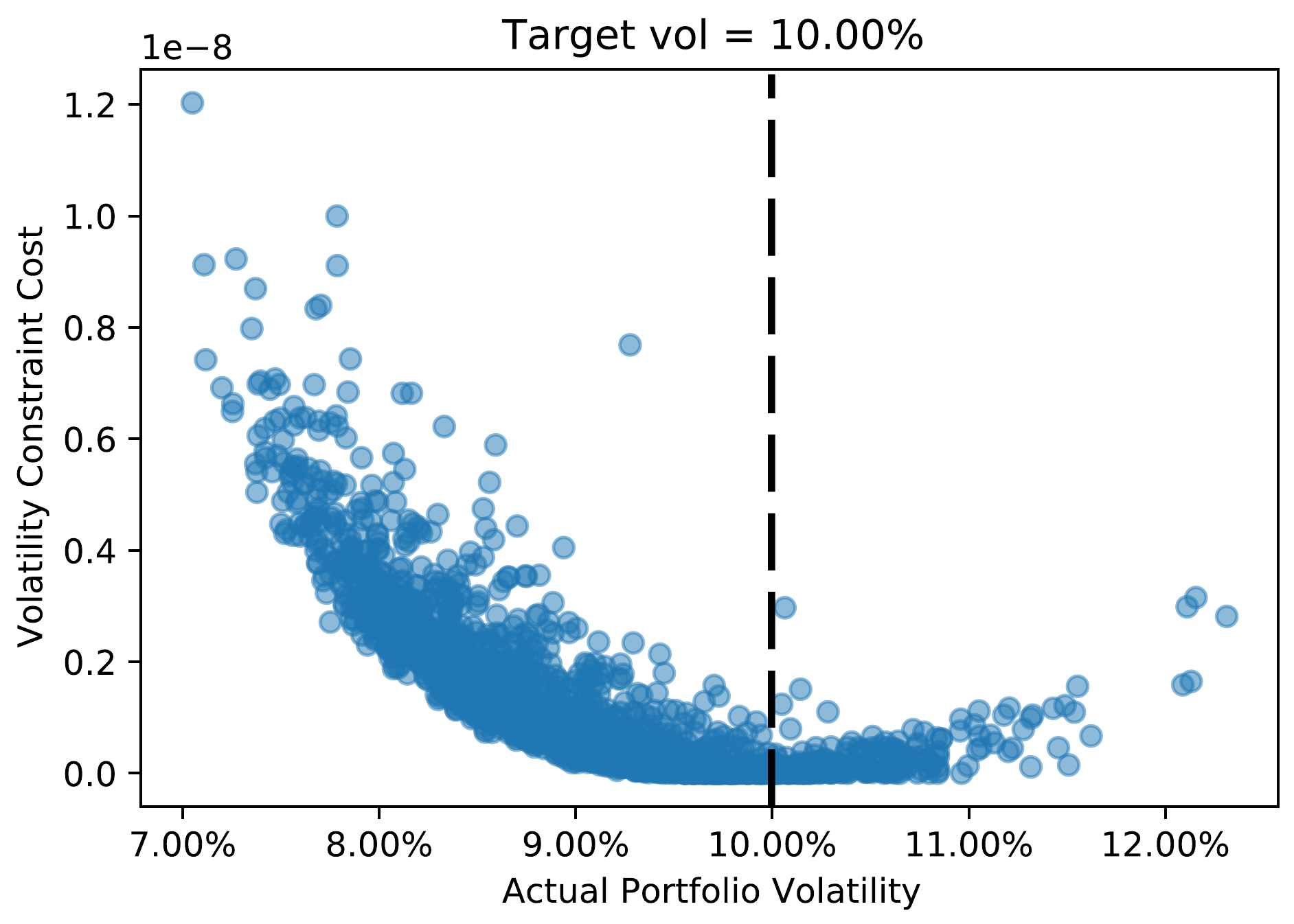}
	\caption{[Color online] Value of the yearly volatility constraint as a function of the yearly portfolio volatility for a fixed target volatility of $10\%$, with $K=100, N_{n,q} = 10$ for all assets $n$. The dotted vertical line is the target volatility of $10\%$.}
	\label{fig1}
\end{figure}

\emph{Investment bands.---} Let us now consider how to include investment bands for each asset $n$. In simple words, this means that there is a minimum and a maximum investment for each asset, which in our language translates to the constraint 
\begin{equation}
\w_n \in \left[ \w^{{\rm min}}_n, \w^{{\rm max}}_n \right], 
\end{equation}
with $ \w^{{\rm min}}_n, \w^{{\rm max}}_n $ being the minimum and maximum (percentual) investments for asset $n$. The reason for this constraint is to impose diversification in the portfolio, so as to avoid possible corner solutions, i.e., portfolios where most of the investment is allocated in very few assets. These solutions, though mathematically correct, may be risky because of uncontrolled reasons not necessarily included in the model. This is the reason why brokers prefer to diversify investments whenever possible for a given return and risk. 

\begin{figure}
	\centering
	\includegraphics[width=0.7\linewidth]{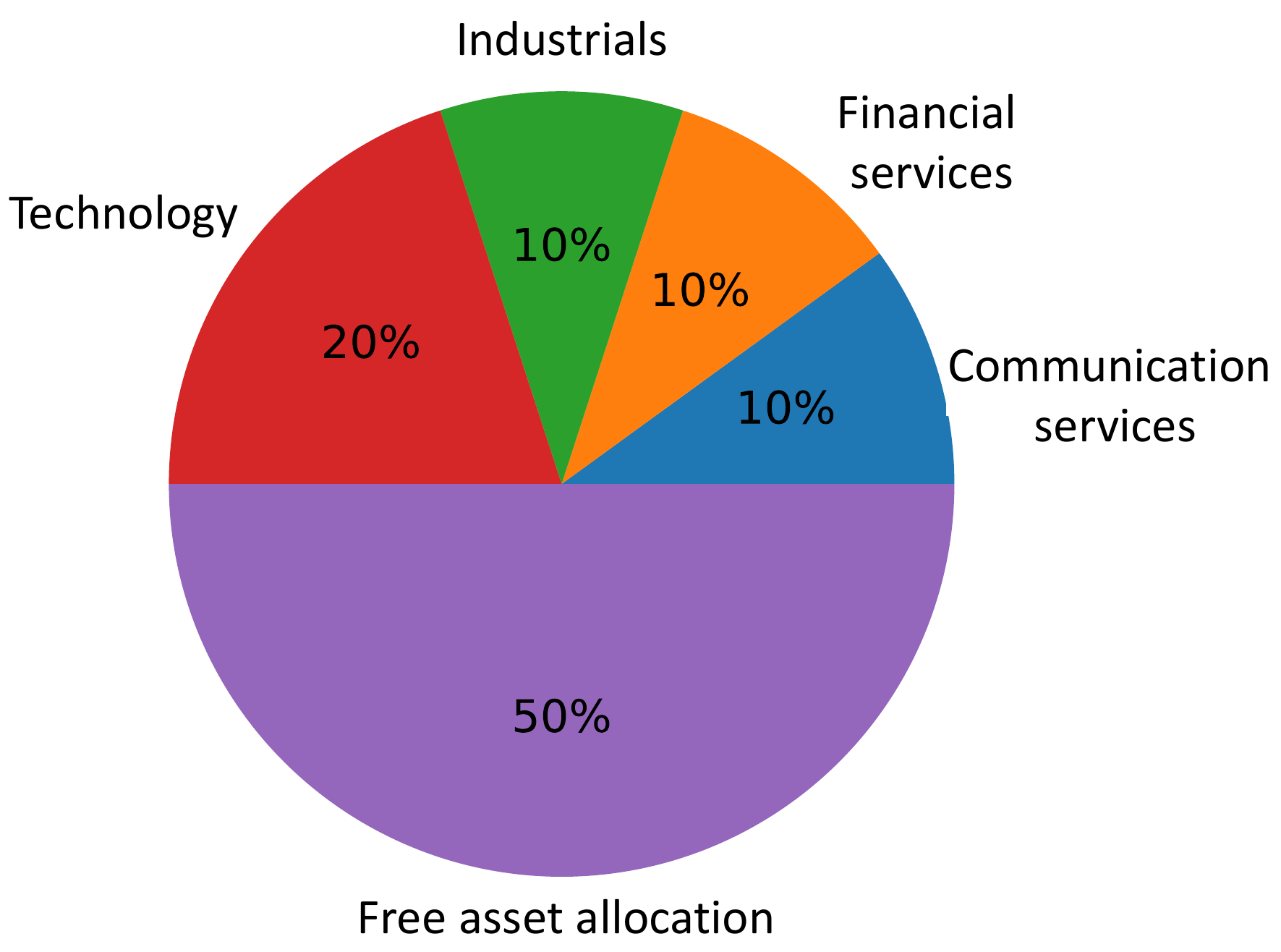}
	\caption{[Color online] Base portfolio composition used in optimizations of the S$\&$P100 and S$\&$P500 in Fig.\ref{fig3}.}
	\label{fig2}
\end{figure}

Imposing these constraints can be done quite naturally in our setting. First, we notice that since $\w_n \ge 0$ always (i.e., we do not allow for \emph{short selling}), then we can consider the following shift for the optimization variables: 
\begin{equation}
\w_n \equiv \left(\w^{{\rm min}}_n + \tilde{\w}_n \right) ~~\tilde{\w}_n \in \left[ 0, \w^{{\rm max}}_n - \w^{{\rm min}}_n \right].
\end{equation}
This linear transformation keeps the optimization problem quadratic as a function of $\tilde{\w}_n$, and  implements the lower cutoff $\w^{{\rm min}}_n$ for each $n$. Next, in order to implement the upper cutoff $\w^{{\rm max}}_n$, we discretize $\tilde{\w}_n$ using bit variables. This will help us in two ways: first, it will impose that investments come in discrete packages, as discussed previously. Second, the fact that we use a fixed number of bits imposes a natural upper cutoff on possible values of the variables. In practice, one can encode $\tilde{\w}_n$ using a binary encoding with $N_{n, q}$ bits in different ways. One that we found useful is the following mapping: 
\begin{equation}
\label{eq:encoding}
\tilde{\w}_{n} = \frac{1}{K} \left( \sum_{q=0}^{N_{n,q}-1}  2^q x_{n, q} + M x_{n, N_{n,q}} \right),
\end{equation}
with $K$ the total money invested, $x_{n, q} \in \{0,1\}$ the readout value of the $q^\text{th}$ bit assigned to asset $n$, and 
$M = K \left(\w^{{\rm max}}_n - \w^{{\rm min}}_n \right)  - \left(2^{N_{n,q}-1} - 1\right)$. In Eq. \eqref{eq:encoding}, the encoding allows for any value of $\w^{{\rm max}}_n$ and $\w^{{\rm min}}_n$ -- though typically both $\w^{{\rm max}}_n$ and $\w^{{\rm min}}_n$ are integers and hence the mapping is \emph{surjective} --, and the \emph{bit depth} $N_{n, q}$ satisfies the constraint 
\begin{equation}
2^{N_{n,q}-1} - 1 \le  K \left(\w^{{\rm max}}_n - \w^{{\rm min}}_n \right), 
\end{equation}
so that $M \ge 0$ always. This procedure naturally implements the diversification constraint individually for each asset, while leaving the overall optimization problem quadratic. It also allows us to fix investment bands on specific sectors by, e.g., distributing equally the minimum and maximum of a sector band amongst all the assets within the sector, which in turn is also a good strategy to avoid corner solutions. 

\begin{figure}
	\centering
	\includegraphics[width=\linewidth]{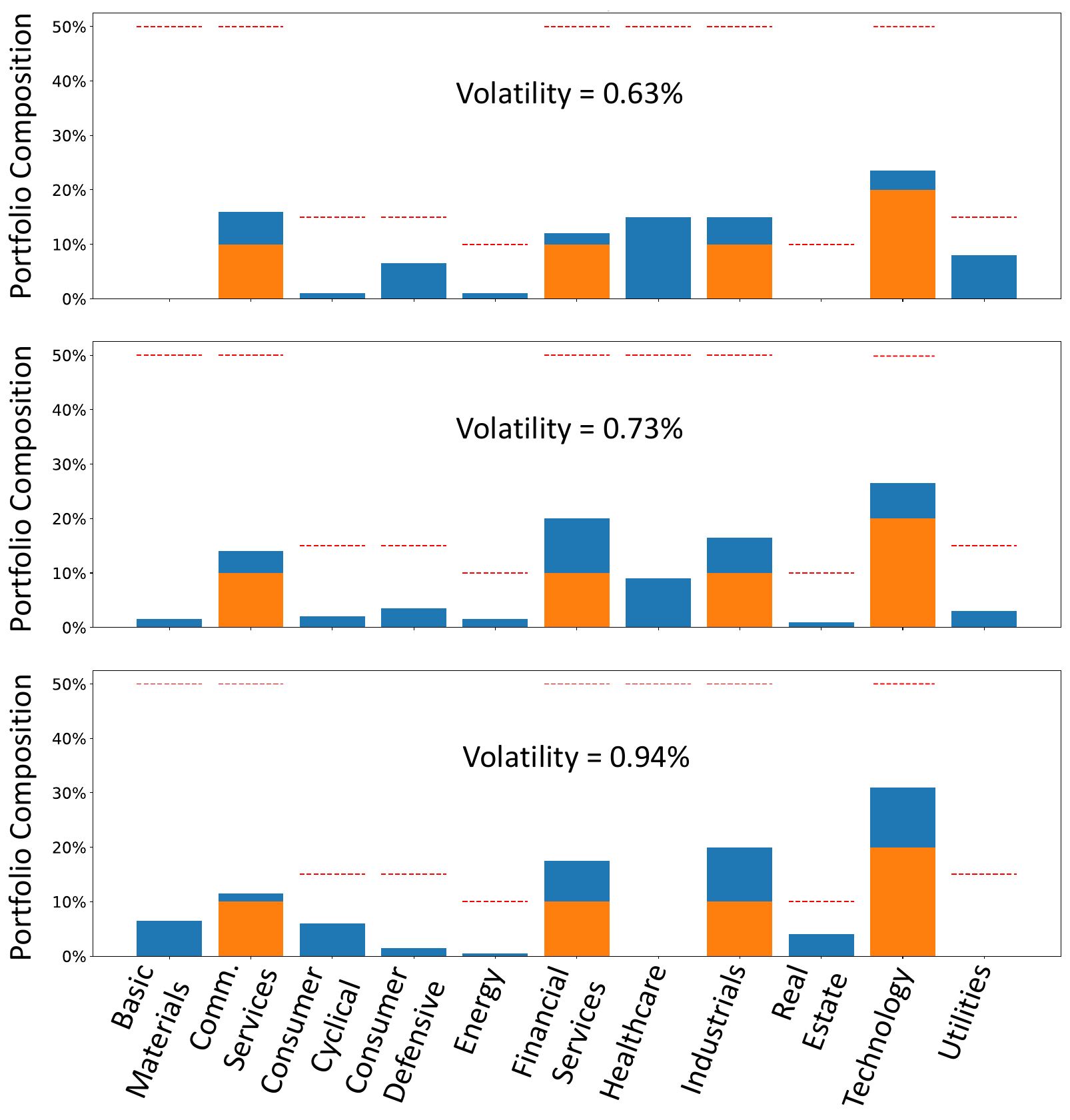}
	\caption{[Color online] Portfolio composition (blue) for investment in the S$\&$P100 with the base from Fig.\ref{fig3} (orange), for different daily volatilities. The maximum investment per sector is shown as an horizontal dashed red line for each sector.}
	\label{fig3}
\end{figure}

\emph{Target volatility.---} In real life settings, financial institutions construct a variety of portfolios for different risk profiles. The problem then is to obtain the best possible portfolio, i.e., the one that maximizes returns, for a given value of the risk as measured by the portfolio's volatility. To do this calculation one could always scan different values of the risk aversion parameter $\gamma$ in Eq.(\ref{eq:bare_cost_function}) so as to obtain different values of the volatility $\sigma$, but this option is clearly not optimal if we are targeting a volatility $\sigma_{{\rm target}}$. To obtain directly the best portfolio with the desired target risk, we can always introduce it as a constraint via a Lagrange multiplier. In such a case, the penalty constraint is 
\begin{equation} 
\mu \left(\w^T \Sigma \w -  \sigma_{{\rm target}}^2 \right)^2, 
\end{equation}
with $\mu$ a Lagrange multiplier, and where we take the squared volatility to avoid square roots.  The problem of this constraint, however, is that it involves up to fourth-order polynomial terms in the final cost function, so that the problem is no longer a QUBO but rather of higher order (HUBO). Solving such problems is known to be quite expensive since they involve order-reduction techniques with an important overhead of bits, see for instance Ref.(\cite{Mugel2020a}). In order to avoid this unpleasant and non-optimal situation, and yet be able to impose the constraint, we \emph{linearize} the portfolio covariance entering the constraint equation. Thus the new constraint is  
\begin{equation} 
\mu \left(k^T \Sigma \w -  \sigma_{{\rm target}}^2 \right)^2, 
\end{equation} 
where $k$ is a vector of constants usually referred to as \emph{linear weights}. The linearization implies that  the constraint remains quadratic, but at the price of having to find $k$ somehow. Here different options are possible. One could for instance optimize $k$ self-consistently, as in some tensor network algorithms  \cite{ORUS2014117}. Another option is to fine-tune $k$ starting from a suitable approximation such as  
\begin{equation}
k_n = \frac{1}{N} ~~ \forall n. 
\end{equation} 
This initial guess is eventually validated by the accuracy of the numerical results, and it can be improved if required. As we shall see in what follows, the difference between the target volatility $\sigma_{{\rm target}}$ and the actual volatility $\sigma$ is very small in our simulations. This approach is therefore a very good approximation especially as we increase the portfolio diversification, i.e., this approximation performs best when we have high amount of diversification as the holdings become more evenly weighted and thus similar to the approximated holdings vector. In this way we can obtain optimal portfolios within a target risk profile while keeping the optimization problem quadratic, and therefore simpler. 

\begin{figure}
	\centering
	\includegraphics[width=\linewidth]{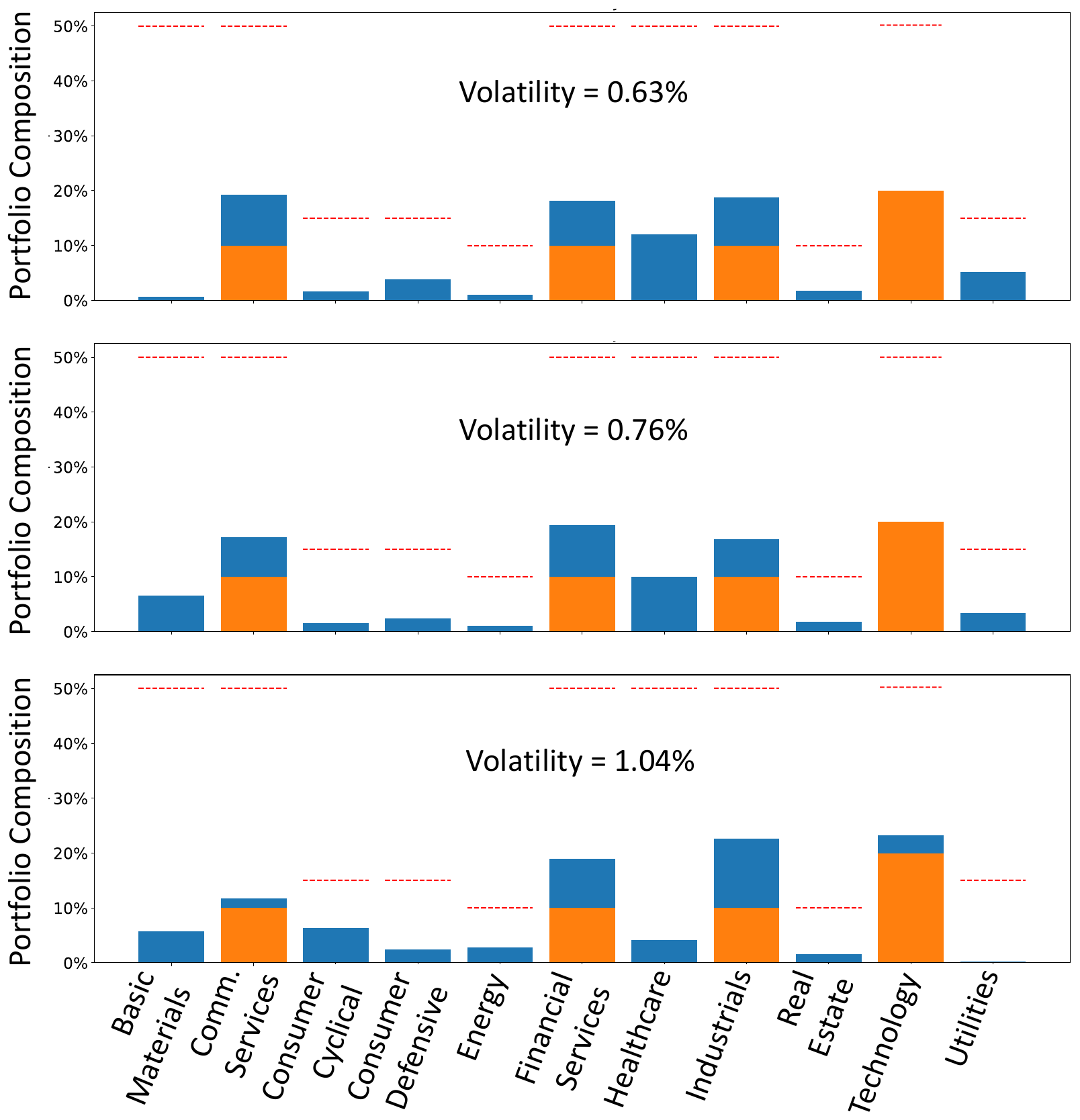}
	\caption{[Color online] Portfolio composition (blue) for investment in the S$\&$P500 with the base from Fig.\ref{fig3} (orange), for different daily volatilities. The maximum investment per sector is shown as an horizontal dashed red line for each sector.}
	\label{fig4}
\end{figure}

\emph{Results.---} To validate our approach we implemented several quantum optimizations of portfolios built by taking all the assets from S$\&$P100 and S$\&$P500. 

The quantum optimizations have been carried out using D-Wave's hybrid quantum annealer with the \emph{Advantage} processor. First, in Fig.\ref{fig1} we show the value of the yearly volatility constraint as a function of the actual portfolio yearly volatility for a portfolio optimization for S$\&$P100 daily closing prices over the year prior to 23-04-2021, and covariance matrix taken over the 3 month period prior to each day. We run these optimizations with parameters such that the portfolio diversification is set to allow a maximum of 10$\%$ of the total portfolio holdings in a single asset. In the figure we can see that the constraint has its minimum essentially at the value of the target volatility, which is around $10\%$. This result shows that the linearization trick can be used to fix a target volatility very efficiently. We also observed that in some optimizations, the introduction of this extra constraint may lead to local minima in the optimizations. This can however be easily handled in several ways, e.g., by playing with the different hyperparameters of the model (such as $\gamma$ and $\mu$), as well as with different initial conditions for the linear weights $k$. 

\begin{figure}
	\centering
	\includegraphics[width=\linewidth]{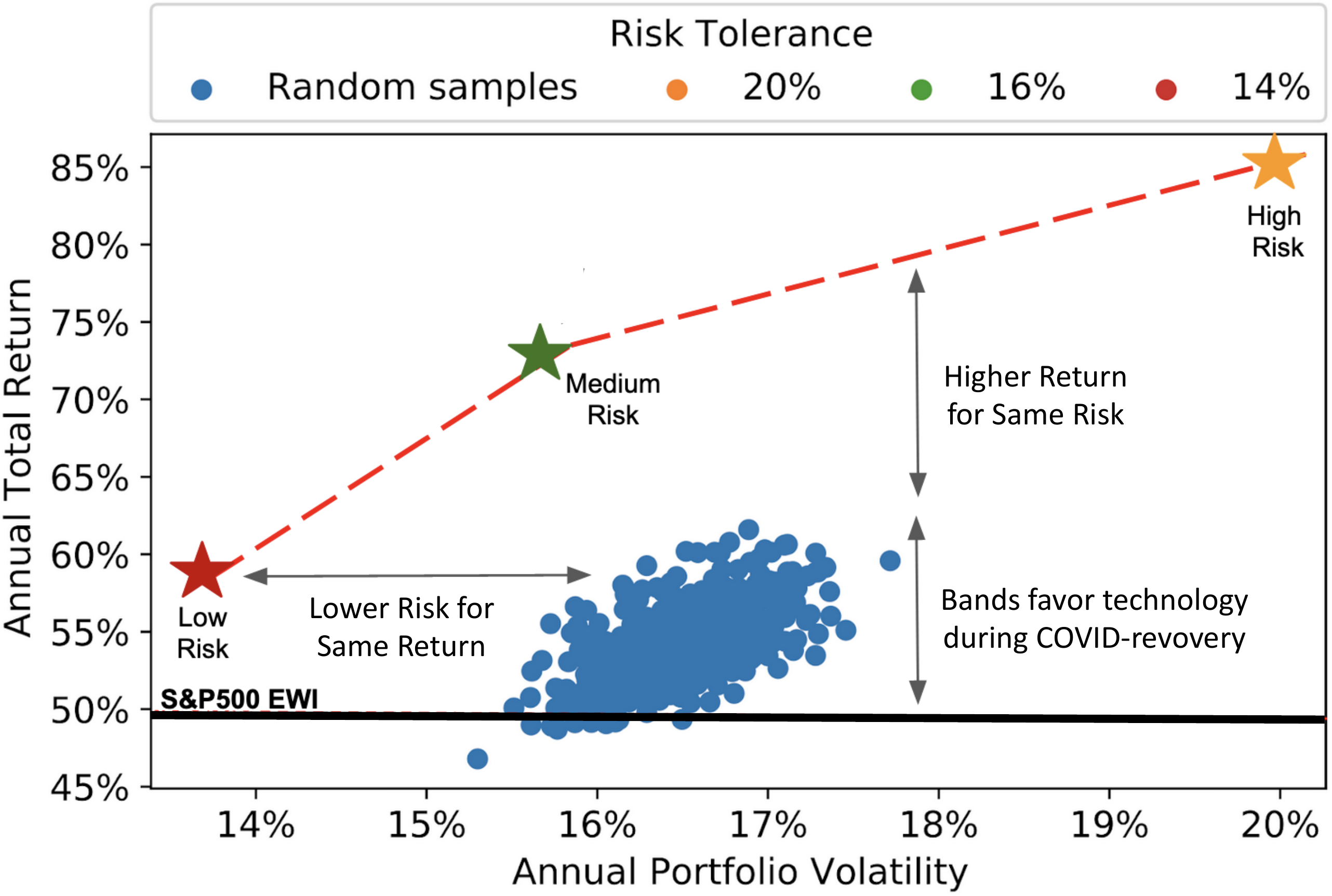}
	\caption{[Color online] Annual return versus annual volatility, for the time period and investment bands specified in the main text. The cloud of blue points are to random typical portfolios. The results from quantum optimization are the stars for low (red), medium (green) and high (yellow) risk profiles. The return of the S$\&$P500 EWI in that period is also shown for reference.}
	\label{fig5}
\end{figure}

Next, we computed the optimal static portfolios (one day) for different target daily volatilities. Following with our analysis, we implemented investment bands on specific sectors of the S$\&$P100 and S$\&$P500 with the base portfolio composition from Fig.\ref{fig2}. This is shown in Figs.\ref{fig3} and \ref{fig4} respectively, where we show the obtained optimal portfolio compositions for different daily volatilities (and where the target volatilities were $0.5\%, 0.75\%$ and $1.00\%$), organized by investment sectors (trading day 23-04-2021, so these are the optimal investments on one single day). In the plots, we show also the base composition as well as the maximum investment per sector, i.e., the   investment bands. As we can see from these figures, the obtained compositions always satisfy the required band constraints along with sufficiently meeting our target volatility requirements, thus proving the validity of our approach. It is remarkable that our approach based on quantum computing is able to optimize such large portfolios of real assets in a remarkable short time - a few minutes per job in the worst case -. 

Finally, in Fig.\ref{fig5} we show the results obtained for yearly portfolio optimizations in terms of total return versus volatility, carried over the full S$\&$P500 in the same period as Fig.\ref{fig1}, and with the investment bands as in Fig.\ref{fig2} but with 15$\%$ minimum investment both in Technology and Industrials. In the figure we show also a cloud of points corresponding to random typical portfolios in the optimization landscape (we have checked that many commercial products, obtained with other classical optimization strategies, are actually not much better than those random points - not shown -.). The results for the quantum optimization of the full  S$\&$P500 with our algorithm corresponds to the stars for low (red), medium (green) and high (yellow) risk profiles. As can be seen from the plot, our method outperforms the typical random portfolio in two ways: for the same return we obtain much lower risk, and for the same risk we obtain much higher returns \footnote{As a remark, notice that the target volatilities differ slightly from the computed ones, due to the large solution space being studied and the approximations mentioned previously. The obtained portfolios are however perfectly valid.}. In the plot we also compare against the returns of investing equally on all assets of S$\&$P500 during that period, which is nothing but the S$\&$P500 Equally Weighted Index (EWI). The S$\&$P500 EWI obtained a total return around 50$\%$ for that period, and was therefore not much better than random portfolios. In fact, we also see that random portfolios tend to be better than equally investing on all assets. This has a simple explanation: the chosen investment bands tend to favor sectors, such as technology, that got very high returns during the COVID-recovery period, coinciding with the time window that we considered.

\emph{Conclusions.---} Here we have shown how the problem of portfolio optimization can be run under realistic conditions on quantum computers. In particular, we have shown how to implement investment bands, as well as how to target specific volatilities in a very efficient way. We have validated our approach by computing optimal portfolios for the full S$\&$P100 and  S$\&$P500, using D-Wave hybrid running on the \emph{Advantage} quantum annealing processor. To the best of our knowledge, these are the largest portfolio optimizations carried on a quantum computer and under real market conditions. We believe that our work will clear the way to quantum computers towards becoming a production-standard tool in quantitative finance. In particular, we believe that using clustering and quantum optimization algorithms together would allow to replicate the behaviour of more complex indexes, such as Nasdaq Composite or others, in turn being particularly useful to build and replicate Exchange Traded Funds (ETF).

{\bf Acknowledgments:} We thank the fantastic technical team at Multiverse, Gianni del Bimbo, Cristina Sanz, Usman Ayub Sheikh, Saeed S. Jahromi, Pablo Martin, and Enrique Lizaso, D-Wave's technical team, and the advise and good guidance of Christophe Jurczak, Pedro Luis Uriarte, Pedro Mu$\tilde{{\rm n}}$oz-Baroja,  Joseba Sagastigordia,  Creative Destruction Lab, BIC-Gipuzkoa, DIPC, Ikerbasque, and Basque Government. 

\bibliography{bibliography}
\bibliographystyle{apsrev4-1}

\end{document}